\documentclass{sig-alternate}
\usepackage{xkeyval}
\usepackage{textcomp}

\begin{document}






%


\title{Analysis of Interference between RDMA and 
Local Access on Hybrid Memory System}

%
%
%
%
%

\numberofauthors{1} 
%
\author{
%
%
\alignauthor
Kazuichi Oe\\
       \affaddr{National Institute of Informatics (NII)}\\
       \email{koe@nii.ac.jp}
}



\maketitle

\begin{abstract}
We can use a hybrid memory system consisting of DRAM and 
Intel\textregistered  
Optane\texttrademark 
DC Persistent Memory (We call it \\
``DCPM'' in this paper) as DCPM 
is now commercially available since April 2019. 
Even if the latency for DCPM is several times higher than that for 
DRAM, the capacity for DCPM is several times higher than that for 
DRAM and the cost of DCPM is also several times lower than that 
for DRAM. 
In addition, DCPM is non-volatile. 
A Server with this hybrid memory system could improve the performance 
for in-memory database systems and virtual machine (VM) systems 
because these systems often consume a large amount of memory. 
Moreover, a high-speed shared storage system can be implemented by 
accessing DCPM via remote direct memory access (RDMA). 
I assume that some of the DCPM is often assigned as a shared area 
among other remote servers because applications executed on a server 
with a hybrid memory system often cannot use the entire capacity of 
DCPM.
This paper evaluates the interference between local memory access and 
RDMA from a remote server. 
As a result, I indicate that the interference on this hybrid memory 
system is significantly different from that on a conventional 
DRAM-only memory system. 
I also believe that some kind of throttling implementation is 
needed when this interference occures.
\end{abstract}

%
%


%
%

%
%




\section{Introduction}

\label{intro}

In recent years, many kinds of persistent memory (PM) 
\cite{DataTiering-EuroSys2016} have been under research 
and development, and some achievements have been merged 
into products for solid state drives (SSDs) and dual 
inline memory modules (DIMMs). 
Intel was commercially available for 
Intel\textregistered  
Optane\texttrademark 
DC Persistent Memory (We call it ``DCPM'' in this paper) 
\cite{INTEL_DCPM} in April of last year. 
DCPM is connected to a computer system via a DIMM slot and 
is available not only for memory but also for storage 
\cite{INTEL_DCPM2,INTEL_DCPM3}. 
DCPM also has a byte-addressable feature, its latency  
is two to five times higher than that of  DRAM's latency
\cite{SWANSON-REPORT,DAMON-19}, 
and its capacity is up to 3 TB per CPU socket. 
For example, a server system that has a two-socket CPU can 
implement a capacity of up to 6 TB for DCPM. 
DCPM must be mounted with DRAM. 
Its capacity is larger than that of DRAM and it is non-volatile. 
A server with DCPM is often used in the operation of an in-memory 
database system or virtual machine (VM) because these applications 
need to use a large amount of memory capacity. 
In particular, there has been much research 
\cite{VLDB-17,SIGMOD-15,SIGMOD-18} on in-memory database 
systems using this hybrid memory system. 
In the VM field, a server can operate a higher number of VMs by 
using the hybrid memory system. 
However, I think that there are many use cases in which a 
server with this hybrid storage system does not consume the entire 
DCPM capacity when executing applications on the server. 
In these cases, we can operate the unused DCPM capacity as 
shared memory or storage among non-DCPM servers. 
A non-DCPM server can execute high-throughput and low-latency 
communication by using remote direct memory access (RDMA), 
which is supported by InfiniBand\texttrademark etc. 
In a word, this hybrid memory system might be accessed from 
local applications and remote applications simultaneously. 
By the way, Imamura et al.\cite{IXPUG-20} reported that the  
interference on this hybrid memory system is significantly 
different from that on a conventional DRAM-only memory 
system when several applications were executed on same server 
simultaneously. 
Therefore, I assume that similar interference will occur 
when the hybrid memory system is accessed from local 
applications and remote applications simultaneously. 
In this paper, I evaluated interference in case that DRAM access 
from a local server and DCPM access from a remote non-DCPM server 
were executed simultaneously. 
I used the Intel\textregistered Memory Latency Checker (
We call it ``MLC'' in this paper)
\cite{INTEL_MLC} 
as the DRAM access application and the ib\_write\_bw, which is 
one of the InfiniBand Verbs Performance Tests\cite{IB_PERFTEST}, 
extended tool as the DCPM access application. 
I called this ib\_write\_bw extended tool ``ib\_write\_bw+''. 
Then, I indicate that the interference on this hybrid memory 
system is significantly different from that on a conventional 
DRAM-only memory system.
I moreover propose that some kind of throttling technique for 
DCPM access from remote non-DCPM server is needed when this 
interference occurs. 

\section{Background}

\label{bak}

\subsection{What is persistent memory (PM)}

\label{bak_pm}

Much research has been done in the persistent memory (PM) field. 
Compared with DRAM, its strong points are that it is low in cost 
and has a large capacity, and its weak point is that its write 
latency is high. Its write latency is about two to five times 
higher than that of DRAM
\cite{DataTiering-EuroSys2016} . 
Consumers can currently use PM because Intel released 
Intel\textregistered  
Optane\texttrademark 
DC Persistent Memory (DCPM) in April 2019. 
Compared with DRAM, DCPM's features are byte-addressable, 
non-volatile, 4 times larger capacity, 2 to 5 times higher write 
latency, and several times lower giga byte costs\cite{IXPUG-20}. 
Consumers can use larger capacity DCPM in the near future because 
Intel\textregistered plans to update the current DCPM. 

\subsection{Hybrid memory system using DCPM and DRAM}

\label{bak_hybrid}

\begin{figure}[tb]
\begin{center}
\includegraphics[width=7.5cm]{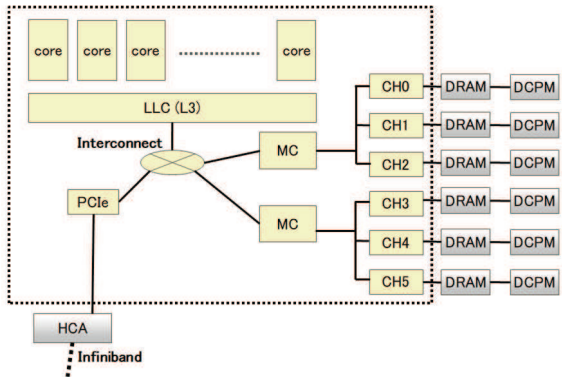}
\caption{Intel CPU configuration with DCPM}
\label{XEON_DCPM}
\end{center}
\vskip-\lastskip \vskip -5pt 
\end{figure}

Figure \ref{XEON_DCPM} shows the configuration of a CPU with DCPM. 
The CPU cores, memory controller (MC), and PCI Express (PCIe) 
are connected by an interconnect. 
The MC has multiple channels (CHs), and each CH is connected to 
both DRAM and DCPM. DCPM must be connected to a CH with DRAM. 
When an application accesses the DCPM area by InfiniBand RDMA, 
the access path for RDMA is from PCIe to MC via the interconnect. 
The path does not include the CPU's last level cache (LLC). 

\subsection{How to access DCPM from application}

\label{bak_app}

A hybrid memory system consisting of DRAM and DCPM needs to set 
either memory mode or app direct mode\cite{INTEL_DCPM2}. 
Memory mode treats DCPM as volatile memory, and DRAM is the cache 
area of DCPM. 
The cache control mechanism is installed on MC. 
App direct mode treats DCPM as non-volatile memory. 
Linux supports three access methods for DCPM : block device, 
filesystem dax, and device dax\cite{COMSYS-19}. 
When using the block device access method, the traditional filesystems 
can be executed on DCPM. But, these filesystems cannot use the 
maximum performance of DCPM because of block unit access. 
Filesystem dax maps the DCPM area to an application's address 
space directly by using dax supported filesystems. 
Device dax maps the DCPM area to an application's address space 
directly by using a device dax driver. 
The device dax is the best method for getting the most out of a 
DCPM's performance. 
Linux also has a patch that treats DCPM as normal RAM
\cite{DCPM-NUMA}. 
The patch can be used to mount the entire DCPM area on a NUMA node 
by using the device dax method, and an application can allocate the 
memory area from the DCPM area by using the Linux numactl command. 

\section{Evaluation}

\label{evo}

\subsection{Overview}

\label{evo_oview}

I wanted to clarify the interference when DRAM or DCPM access from 
a local server and DCPM access from a remote non-DCPM server are 
executed simultaneously. 
The application for the local server was the Intel\textregistered 
Memory Latency Checker (Intel MLC), and the application for remote 
non-DCPM server was the extended ib\_write\_bw. 
As mentioned the extended ib\_write\_bw is called ``ib\_write\_bw+''. 
The evaluation considerd MLC-only performance, ib\_write\_bw+-only 
performance, and the performance when executing MLC and ib\_write\_bw+ 
simultaneously. 
The evaluation also used the Platform Profiler feature of 
Intel\textregistered  VTune Amplifier 2019\cite{INTEL-VTUNE} to 
clarify the internal throughput and latency for both DCPM and DRAM. 

\subsection{Environment}

\label{evo_env}

\subsubsection{System configuration}

\label{evo_env_conf}

\begin{figure}[tb]
\begin{center}
\includegraphics[width=8cm]{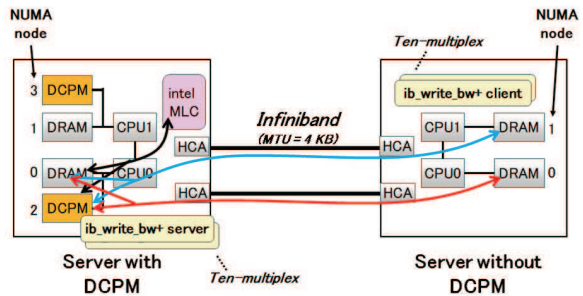}
\caption{Evaluation system}
\label{EX_SYSTEM}
\end{center}
\vskip-\lastskip \vskip -5pt 
\end{figure}

Figure \ref{EX_SYSTEM} shows the evaluation system. 
A server with DCPM and a server without DCPM were connected 
by two InfiniBand paths. 
The server with DCPM consisted of a 16-core Xeon Gold 5218 (Cascade 
Lake) x2, 192 GB of DRAM, and 812 GB of DCPM (256 GB x 6). 
I also set six DCPM DIMMs as one DCPM area by using interleaved 
app direct mode. 
The server was also installed with the NUMA node patch described 
in Section \ref{bak_app}. 
Then, NUMA node 0 and 1 were mounted to DRAM, and NUMA 2 and 3 were 
mounted to DCPM. 
SUSE Linux Enterprise 15 (5.0.0-rc1-25.25) was also installed on 
the server. 
The server without DCPM consisted of a 16-core Xeon E5-2650L (Sandy 
Bridge) x2 and 32 GB of DRAM. 
Fedora30 (5.1.12-300.fc30) was also installed. 
Two InfiniBand Host Channel Adapters (HCA) were also installed on 
both servers, and these servers were connected directly by using the 
InfiniBand. 
The HCA's bandwidth is 100 Gbps per direction, and teh total bandwidth 
is 200 Gbps per direction. 

\subsubsection{Intel MLC}

\label{evo_env_mlc}

In this paper, MLC version 3.7 was used, and the hybrid memory system 
was evaluated by using loaded\_latency mode. 
Its read/write option was R (Read only), W2 (2:1 read-write ratio), 
and W5 (1:1 read-write ratio). 
Its memory size was 4 GB, so the effect of CPU cache can 
be ignored. 
Its offset setting was random (rand) and sequential (seq), and 
its NUMA node was 0 (DRAM) or 2 (DCPM).

\subsubsection{ib\_write\_bw+}

\label{evo_env_ib}

I downloaded and investigated the source code for InfiniBand Verbs 
Performance Tests 3.0 (March 2015). 
In particular, I carefully investigated the ib\_write\_bw source 
code which included the RDMA test. 
Then, the ib\_write\_bw repeatedly executed RDMA with the same source 
and destination address, and no tool existed for the read/write 
mixed RDMA test. 
Most real applications using RDMA often access 
various source/destination addresses, and their operations 
were mostly read-write mixed.
Then, I added the following features to the ib\_write\_bw. 

\begin{itemize}
\item Setting any size for RDMA buffer (in this evaluation, 
10 MB was set). 
\item Choosing three operations [Read only, Write only, 
Read/Write mixed (1:1)].
\item Choosing two RDMA offset updates (random, sequential). 
The offset is updated in the RDMA buffer. 
\end{itemize}

In the evaluation, ten-multiplex ib\_write\_bw+ was used 
(See Figure \ref{EX_SYSTEM}). 
I decided on this in a preliminary experiment to 
create sufficient memory access for DCPM. 

\subsection{How to evaluate hybrid memory system}

\label{evo_howto}

To understand the performance without interference, both 
ib\_write\_bw+-only and MLC-only were evaluated. 
Then, the interference performance when co-executing ib\_write\_bw+ 
and MLC was evaluated. 
I can understand the performance degradation by comparing 
its interference performance and its unit performance. 
Moreover, to find which conditions increased the interference, 
both ib\_write\_bw+ and MLC were executed simultaneously while 
changing their parameters. 

\subsection{Results}

\label{evo_results}

\subsubsection{ib\_write\_bw+ only}

\label{evo_results_rdma}

\begin{figure}[tb]
\begin{center}
\includegraphics[width=7.5cm]{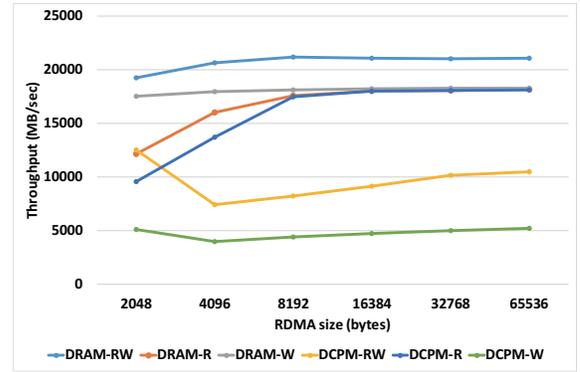}
\caption{RDMA random throughput (MB/sec)}
\label{RDMA_R}
\end{center}
\vskip-\lastskip \vskip -5pt 
\end{figure}

\begin{figure}[tb]
\begin{center}
\includegraphics[width=7.5cm]{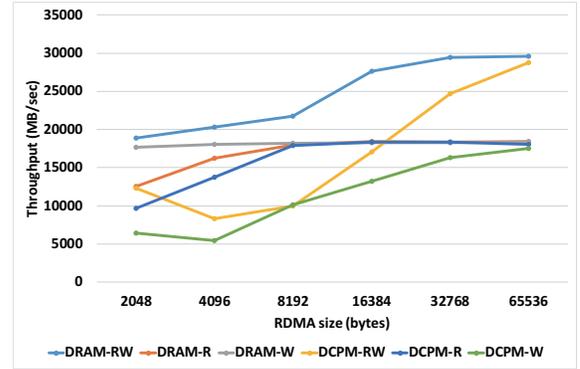}
\caption{RDMA sequential throughput (MB/sec)}
\label{RDMA_S}
\end{center}
\vskip-\lastskip \vskip -5pt 
\end{figure}

The ib\_write\_bw+ server was executed on the server with DCPM and 
the ib\_write\_bw+ client was executed on the server without DCPM. 
The RDMA buffer for the server was set on both DRAM and DCPM. 
In order to generate enough IO traffic, ten-multiplex execution was 
done. 
Figure \ref{RDMA_R} shows the results for a random RDMA offset, and 
Figure \ref{RDMA_S} shows those for a sequential RDMA offset. 
The RDMA size was changed from 2 KB to 64 KB, and the RDMA operation 
was Read, Write, and Read/Write mixed. 
First, the results for the random RDMA offset are discussed. 
When executing RDMA Read, there was near throughput even when 
the server's RDMA buffer was changed from DRAM to DCPM. 
In particular, there was almost the same throughput when the RDMA size 
was more than 8 KB. 
When executing RDMA Write and RDMA Read/Write mixed, the throughput 
setting the server's RDMA buffer to DRAM was three times higher 
than that setting the server's RDMA buffer to DCPM. 
This is because of the higher write latency for DCPM. 
Second, the results for the sequential RDMA offset are discuessed. 
The results for RDMA Read were similar to the results for the random 
RDMA offset. 
However, the results for RDMA Write and Read/Write mixed were 
different from those for the random offset. 
Executing RDMA Write, the difference between the DRAM and DCPM 
throughput decreased when a bigger RDMA size was used. 
Both throughputs matched at an RDMA size if 64 KB. 
These results may have an effect on write buffer of MC. 
Executing RDMA Read/Write mixed, both throughputs reached 30 GB/sec 
when the RDMA size 64 KB. 
This is because the total bidirectional bandwidth reached 400 Gbps. 

\subsubsection{MLC only}

\label{evo_results_mlc}

Table \ref{EVO_MLC_ONLY} shows the results described in Section 
\ref{evo_env_mlc}. 
Thers was one thread execution for MLC. 
The results indicate that the MLC throughputs with DRAM were three to 
four times higher than those with DCPM. 

\begin{table}[tb] \caption{Throughput for MLC only (MB/sec)} 
\label{EVO_MLC_ONLY}
\begin{center}
\begin{tabular}{|l|l|l|l|}
\hline
            & R     & W2    & W5   \\ \hline
rand+DCPM   & 2632  & 2592  & 2326 \\ \hline
seq+DCPM    & 4042  & 5247  & 5715 \\ \hline
rand+DRAM   & 6733  & 8576  & 7758 \\ \hline
seq+DRAM    & 12414 & 17213 & 19943 \\ \hline
\end{tabular}
\end{center}
\end{table}

\subsubsection{Interference between ib\_write\_bw+ and MLC}

\label{evo_results_mlc_rdma}

\paragraph{MLC performance}

\label{evo_results_mlc_rdma1}

\begin{figure}[tb]
\begin{center}
\includegraphics[width=7.5cm]{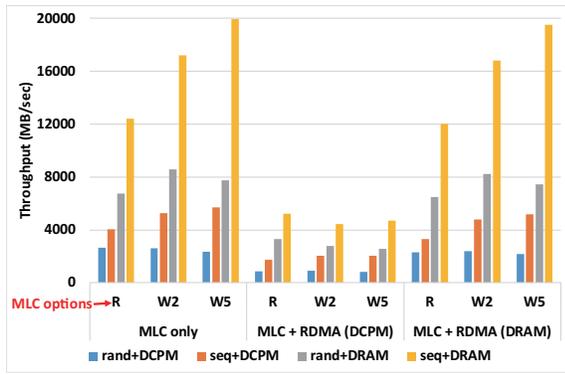}
\caption{MLC results when executing ib\_write\_bw+ 4KB, random offset,
  RW mixed (MB/sec)}
\label{MLC_with_RDMA_1}
\end{center}
\vskip-\lastskip \vskip -5pt 
\end{figure}

Figure \ref{MLC_with_RDMA_1} shows the MLC throughput when 
ib\_write\_bw+ and MLC were co-executed. 
The X-axis indicates the conditions for these evaluations. 
R, W2, and W5 are the MLC options described in Section 
\ref{evo_env_mlc}. 
The server's memory area for ib\_write\_bw+ was both DRAM and DCPM, 
and its other options were 4-KB RDMA size, random offset, Read/Write 
mixed operation, and ten-multiplex execution because these options 
are the condition that generates the most IO accesses on DCPM. 
The Y-axis is the MLC throughput, and the usage guide shows the 
remaining MLC options described in Section \ref{evo_env_mlc}. 
The results shows that the interference performances were less than 
half of the non-interference performances even when the MLC was 
executed with DRAM (See the portion of ``MLC + RDMA(DCPM)''). 
In particular, when the MLC options were W5+seq+DCPM, the MLC throughput 
became 24\% of the MLC only throughput. 
I guess that the MLC throughput of DRAM became drastically slow down 
because both DRAM and DCPM were connected to the same CH. 
If many IO accesses are concentrated to DCPM, IO accesses for DRAM 
may be waited till the completion of DCPM accesses 
(See figure \ref{XEON_DCPM}). 
However, when the server's memory area for ib\_write\_bw+ was DRAM, 
small throughput falls occurred (lowered less than 20\%) 
(See the portion of ``MLC + RDMA(DRAM)''). 

\paragraph{ib\_write\_bw+ performance}

\label{evo_results_mlc_rdma2}

\begin{figure}[tb]
\begin{center}
\includegraphics[width=7.5cm]{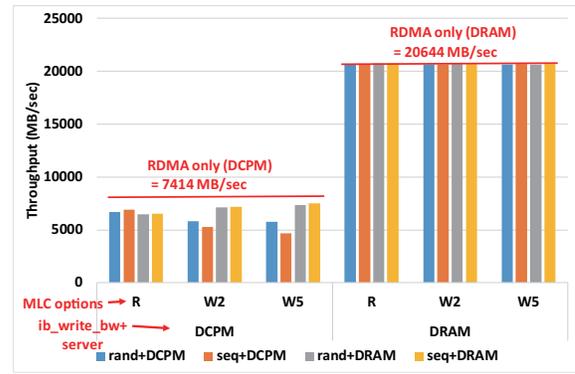}
\caption{ib\_write\_bw+ results when executing MLC (MB/sec)}
\label{MLC_with_RDMA_2}
\end{center}
\vskip-\lastskip \vskip -5pt 
\end{figure}

Figure \ref{MLC_with_RDMA_2} shows the ib\_write\_bw+ throughput when \\
ib\_write\_bw+ and MLC were co-executed. 
Both the X-axis and the usage guide are almost the same as Figure 
\ref{MLC_with_RDMA_1}. 
The Y-axis is the throughput for ten-multiplex ib\_write\_bw+ 
executions. 
The results shows less than 20\% degradation when the server's memory 
area for ib\_write\_bw+ was DCPM. 
Moreover, tiny throughput falls occurred when the server's memory 
area for ib\_write\_bw+ was DRAM. 
The previous paragraph showed that the MLC performance was 
drastically slow down when the ib\_write\_bw+ was executed to 
DCPM. 
However, the ib\_write\_bw+ performance was tiny slow down even if 
the MLC target was DCPM. 
I guess that the implementation for Intel's MC brings this 
phenomenon. 

\paragraph{MLC performance when traffic for ib\_write\_bw+ was changed}

\label{evo_results_mlc_rdma3}

\begin{figure}[tb]
\begin{center}
\includegraphics[width=7.5cm]{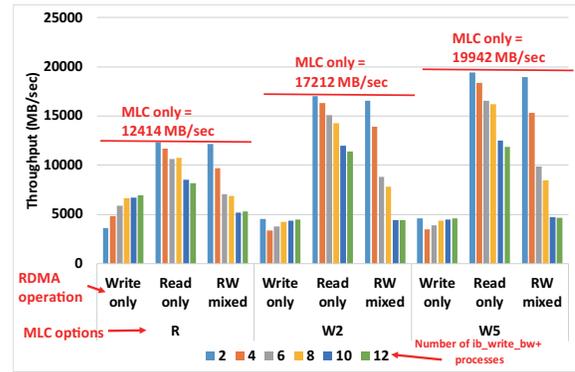}
\caption{MLC results with various RDMA settings (MB/sec)}
\label{MLC_with_RDMA_3}
\end{center}
\vskip-\lastskip \vskip -5pt 
\end{figure}

\begin{figure}[tb]
\begin{center}
\includegraphics[width=7.5cm]{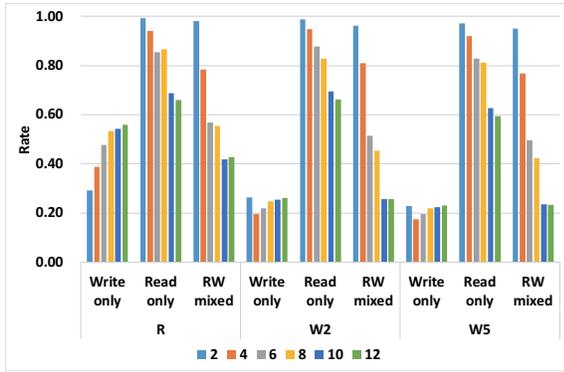}
\caption{MLC results with various RDMA settings (rate)}
\label{MLC_with_RDMA_4}
\end{center}
\vskip-\lastskip \vskip -5pt 
\end{figure}

\begin{figure}[tb]
\begin{center}
\includegraphics[width=7.5cm]{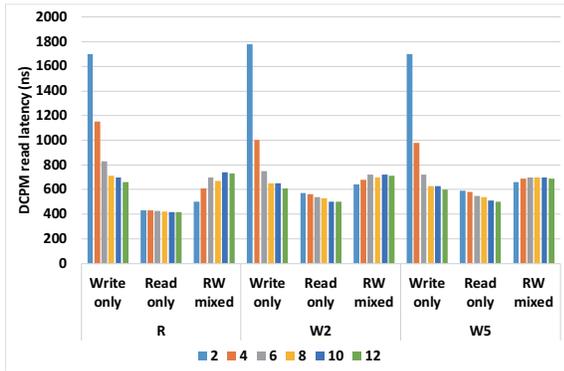}
\caption{DCPM read latency when using Intel VTune (ns)}
\label{MLC_with_RDMA_5}
\end{center}
\vskip-\lastskip \vskip -5pt 
\end{figure}

\begin{figure}[tb]
\begin{center}
\includegraphics[width=7.5cm]{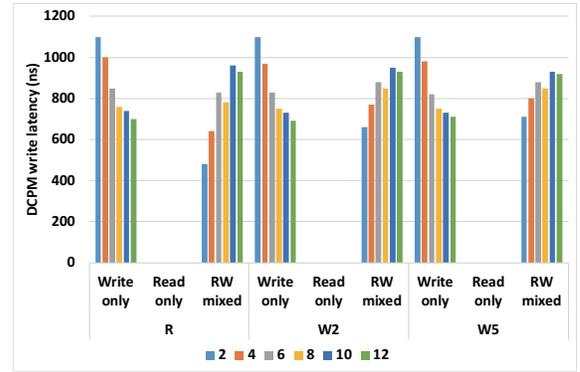}
\caption{DCPM write latency when using Intel VTune (ns)}
\label{MLC_with_RDMA_6}
\end{center}
\vskip-\lastskip \vskip -5pt 
\end{figure}

From the results so far, this paper has indidated that MLC throughput 
is drastically changed when ib\_write\_bw+ using DCPM and MLC are 
co-executed. 
I also investigated the MLC throughput when both the amount of 
RDMA access for DCPM and the RDMA access patterns were changed. 
Both can be adjusted by changing the parameters for \\
ib\_write\_bw+. 
In particular, the RDMA operations were not only RDMA Read/Write mixed 
but also Read-only and Write-only. 
The multiplex values for ib\_write\_bw+ ranged from 2 to 12 so as to 
change the amount of RDMA access. 
The MLC operations were R, W2, and W5 when the offset was sequential 
to DRAM. 
Figure \ref{MLC_with_RDMA_3} shows the results. 
The usage guide indicates the multiplex value for ib\_write\_bw+. 
All of the results (R, W2, and W5) indicate that RDMA Write only 
caused the interference to be big even when the multiplex value was 
small. 
However, both RDMA Read-only and RDMA Read/Write mixed increased 
the interference as the multiplex value increased. 
Figure \ref{MLC_with_RDMA_4} indicates the rate for each result 
when the result of MLC-only was 1.00. 
It can be seen that interference was big for RDMA Write-only. 
In particular, the MLC throughput co-executing with ib\_write\_bw+ 
was 18\% of MLC-only when the ib\_write\_bw+ options were four 
multiplexes and RDMA Write-only and the MLC option was W5. 
To determine the reason for Figure \ref{MLC_with_RDMA_3} and 
\ref{MLC_with_RDMA_4}, the read and write latency were also 
investigated by using the Platform Profiler feature of 
Intel\textregistered 
VTune Amplifier 2019. 
Figure \ref{MLC_with_RDMA_5} shows the results for read latency, 
and Figure \ref{MLC_with_RDMA_6} shows those for write latency. 
First, RDMA Write-only is discussed. 
Both read and write latencies were higher when the multiplex value
was smaller as can be seen from the results of both figures. 
The MLC throughput slowed down when these latencies 
increased. 
%

%
Next is RDMA Read-only. 
The read latencies were stable when the MLC was executed with the R 
option, and the read latencies became a little lower when the multiplex 
value was bigger with the W2 and W5 options.
The amount of RDMA read accesses was bigger when the multiplex value
was bigger.
Then, the increases for RDMA read accesses afforded the large 
interference. 
Last is RDMA Read/Write mixed. 
Both read and write latencies were higher when the multiplex value
was bigger, and the amount of RDMA read/write accesses was bigger 
when the multiplex value was bigger. 
This is why the large interference for MLC throughput. 

\section{Discussion}

\label{discussion}

From the results of Section \ref{evo_results}, the MLC throughput 
drastically changed when both ib\_write\_bw+ for the DCPM area 
and MLC for the DRAM/DCPM area were executed simultaneously. 
However, the ib\_write\_bw+ throughput changed a little 
(up to 20\%). 
Then, I proposed that some kind of throttling technique for the 
DCPM access from remote non-DCPM server (ib\_write\_bw+ in this 
paper) was needed when this interference occurred. 
For example, when a program executes many RDMA requests for DCPM, 
it checks the amount of MC accesses by using the CPU's statistical 
function. 
If the amount of MC access is bigger, the program should reduce the 
number of RDMA requests until the interference is not occurred. 
Figure \ref{MLC_with_RDMA_4} indicated that the interferences for 
Read-only and Read/Write mixed RDMA were not occurred when the 
number of ib\_write\_bw+ process was smaller. 
When Write-only RDMA, I confirmed that the interference was not 
occurred when the ib\_write\_bw+ was executed by using one HCA 
pair only. 
I will study the throttling technique in the future work. 
Imamura et al.\cite{IXPUG-20} reported that the interference 
on this hybrid memory system is significantly different from 
that on a conventional DRAM-only memory system when several 
applications were executed on same server simultaneously. 
He also indicated that DRAM access may be changed because the 
write queue in MC keeps a large number of write requests. 
Moreover, DRAM access starts to be changed when the write 
latency for DCPM reaches 1.5 micro seconds. 
The results for this paper also indicated the interference 
between DCPM accessing application when useing RDMA and DRAM 
accessing application when using MLC. 
However, from this paper's results, the DRAM access started to 
be changed when the write latency for DCPM was from 0.7 to 0.9 
micro seconds.
Both the MLC and RDMA execution shared the resources from MC to 
DRAM or DCPM. 
Therefore, the MLC throughput changed because 
some resource competition for MC occurred. 
The resource competition points may include a new point other 
than Imamura's report because of the difference in the write 
latency for DCPM. 

\section{Related work}

Many research papers already have evaluated DCPM. 
Izraelevitz et al.\cite{SWANSON-REPORT} evaluated each DCPM 
function exhaustively. 
Renon et al.\cite{DAMON-19} executed a basic evaluation for 
applying database logging. 
Hirofuchi et al.\cite{HIROFUCHI} executed an evaluation for 
applying virtual machines (VMs).
Weiland et al. \cite{DCPM-SC19} executed an evaluation for 
high-performance scientific applications. 
These pieces of research indicated that the latency for DCPM changed 
between 100 ns and 800 ns according to their access pattern for 
DCPM. 

\section{Conclusion}

This paper evaluated interference for when DRAM access from a 
local server and DCPM access from a remote non-DCPM server 
was executed simultaneously. 
An Intel\textregistered  Memory Latency Checker (Intel MLC) was 
used as the DRAM access application, and the the ib\_write\_bw, 
an InfiniBand Verbs Performance Test, extended tool was used 
as the DCPM access application, called ``ib\_write\_bw+''. 
This paper showed the interference on this hybrid memory 
system is significantly different from that on a conventional 
DRAM-only memory system.
I moreover propose that some kind of throttling technique for 
DCPM access from remote non-DCPM server is needed when this 
interference occurs. 
I would like to study the throttling technique in the future. 

\newpage

%
\bibliographystyle{abbrv}
\bibliography{dcpm_rdma_evo_preprint}  
%
%

\end{document}